\documentclass[3p,twocolumn,times,authoryear,12pt]{elsarticle}

\usepackage{amssymb}
\journal{Icarus}
\usepackage{mathtools, cuted}
\usepackage{kpfonts}
\usepackage[T1]{fontenc}
\usepackage{lettrine}
\usepackage{xcolor}
\usepackage[citebordercolor=white]{hyperref}

\begin{document}

\begin{frontmatter}

\title{\raggedright \huge {\color{black}Saltation under Martian Gravity and its Influence on the Global Dust Distribution}}

\author[label1]{Grzegorz Musiolik}
\ead{gregor.musiolik@uni-due.de}
\author[label1]{Maximilian Kruss}
\author[label1]{Tunahan Demirci}
\author[label1]{Bj\"orn Schrinski}
\author[label1]{Jens Teiser}
\author[label2]{Frank Daerden}
\author[label3]{Michael D. Smith}
\author[label2]{Lori Neary}
\author[label1]{Gerhard Wurm}

\address[label1]{University of Duisburg-Essen, Faculty of Physics, Lotharstr. 1-21, 47057 Duisburg, Germany}
\address[label2]{Royal Belgian Institute for Space Aeronomy (BIRA-IASB), Ringlaan 3, B-1180 Brussels, Belgium}
\address[label3]{NASA Goddard Space Flight Center, Greenbelt, MD 20771, United States}

\begin{abstract}
	Dust and sand motion are a common sight on Mars. Understanding the interaction of atmosphere and Martian soil is fundamental to describe the planet's weather, climate and surface morphology.
	
	We set up a wind tunnel to study the lift of a mixture between very fine sand and dust in a Mars simulant soil. The experiments were carried out under Martian gravity in a parabolic flight. The reduced gravity was provided by a centrifuge under external microgravity. The onset of saltation was measured for a fluid threshold shear velocity of 0.82$\pm$0.04 m/s. This is considerably lower than found under Earth gravity.
	
	In addition to a reduction in weight, this low threshold can be attributed to gravity dependent cohesive forces within the sand bed, which drop by 2/3 under Martian gravity. The new threshold for saltation leads to a simulation of the annual dust cycle with a Mars GCM that is in agreement with observations.
	
\end{abstract}


\end{frontmatter}


\section{Introduction}

\lettrine{W}{ind} tunnel experiments simulating dust lifting on the Martian surface date back into the last century \citep{Greeley1980}. These studies use different low-density materials to simulate the reduced gravity on Mars of 0.38 g and provide the first thresholds for the onset of saltation. Compared to the available meteorological data which allows an estimation of the Martian boundary layer winds \citep{Hess1977, Schofield1997, Magalhaes1999, Holstein2010} and to predictions from global circulation models (GCMs) \citep{Forget1999,Haberle1999,Haberle2003}, this threshold should be exceeded only rarely \citep{Jerolmack2006,Kok2012,Wang2015,Newman2017}. In contradiction to this, the motion of dust and sand can be observed frequently and has a large impact on the Martian climate \citep{Zurek1992,Smith2004,Heavens2011,Guzewich2017}.

Strong efforts have been made in recent years to detail the picture of soil-atmosphere interaction \citep{White1987, Strausberg2001, Sullivan2005, Greeley2006, Merrison2007, Almeida2008, Merrison2008, Sullivan2008, Kok2010a, Kok2010b, Bridges2012}. Even though, it still remains questionable if dust storms can generally be initiated by wind drag. For example, a lower shear velocity would suffice to keep saltation active but cannot explain the onset of saltation. Hence, also supporting effects are studied. For example,  insolation of the soil leads to thermal creep and a sub-surface overpressure, capable of reducing the threshold wind velocity significantly \citep{deBeule2014, Kuepper2015}. Also dust devils go along with pressure excursions which can support grain lifting \citep{Balme2006}. In any case, numerical models often use an artificially reduced threshold which is needed to initiate lifting events to simulate saltation on Mars \citep{Haberle2003,Kahre2006,Daerden2015}.  

However, aeolian transport experiments at Martian gravity and pressure, as e.g. by \citet{White1987}, are rare. In this work, we investigate the influence of reduced gravity on saltation and show that 
the threshold velocity for a sand bed prepared  and subject to gas flow at Martian gravity and pressure is strongly reduced.

\subsection{Experimental setup}

\begin{figure*}[h!]
	\centering
	\includegraphics[width=0.8\textwidth]{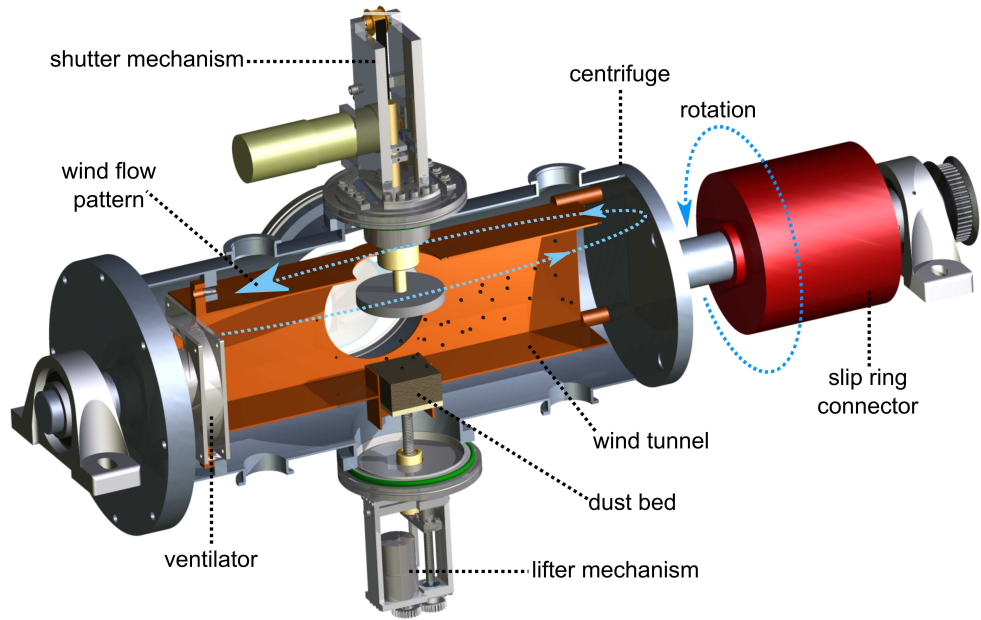}
	\caption{\label{fig.exp}A schematics of the experiment. The outer vacuum chamber has a diameter of 200 mm and is designed as a centrifuge. The wind tunnel is placed inside this centrifuge and has a cross section of 100 mm $\times$ 100 mm.}
\end{figure*}

The Martian environment is simulated in a low pressure wind tunnel designed simultaneously as a centrifuge to simulate Martian conditions (fig. \ref{fig.exp}). In detail, the experiment consists of a vacuum chamber which is evacuated to a pressure of 6 mbar and a gas mixture of 95\% CO$_2$ and 5\% air. It has a radius of 100 mm and can be rotated at more than 2 Hz. The wind tunnel is located in the center of the experiment chamber and has a cross section of 100 mm $\times$ 100 mm. The wind flow is created by a fan rotating with up to 11.000 rpm at an air flow rate of up to 570 m$^3$/h. The gas flows through the wind tunnel over the sand bed and back again on the outer side of the wind tunnel. The total mass of the experiment is 161 kg. The Reynolds number for this configuration inside the wind tunnel is on the order of $Re \approx 800$. 

\begin{figure}[h!]
	\centering
	\includegraphics[width=0.49\textwidth]{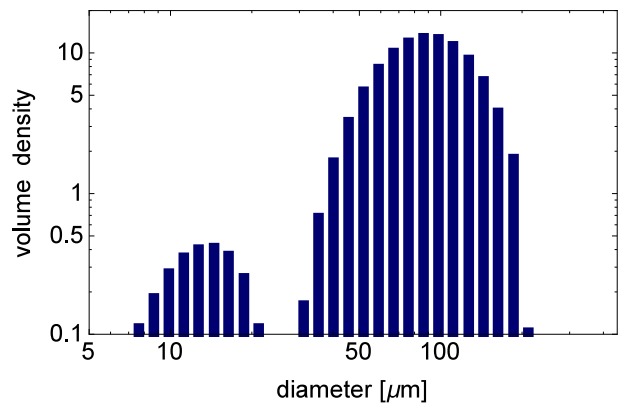}
	\caption{\label{fig.size}Grain size distribution of the used sample. The fraction of larger grains dominates the mass distribution and therefore the mechanical properties of the sample.}
\end{figure}

The set up is used in parabolic flights on the ZERO-G Airbus operated by NOVESPACE in Bordeaux \citep{Pletser2016}. A flight consists of 31 parabolas with a duration of 22 s per parabola and a residual acceleration on the scale of $\pm$0.05 g \citep{Pletser2016}. The centrifugal force on the surface of the dust bed is set to 0.38 g, while additional experiments on ground were carried out at 1 g. The particle sample was $\sim$ 50 g of a mixture between very fine sand and dust consisting of the JSC 1A Martian regolith simulant, which was tempered at 600 K before to remove volatiles and organics. This simulant is made out of altered volcanic ash from a Hawaiian cinder cone and is a representative species for the reflectance spectrum, mineralogy, chemical composition, density, porosity and magnetic properties of the Martian soil \citep{Allen1997}. The size distribution of the used sample is shown in fig. \ref{fig.size}.

Before each parabola, the sample is closed by a shutter mechanism to protect the sample against uncontrolled accelerations. The experiment runs automatically. With the onset of the microgravity phase, the chamber starts to rotate. The shutter is removed once the set rotation frequency is established. Due to the momentum of the shutter, the sand sample is first lifted and then settles back to the ground. This way, the surface of the sand sample is prepared at Martian gravity level before each measurement.
The erosion is observed optically with a camera installed perpendicular to the wind flow at 457 frames per second and an exposure time of 200 $\mu$s, using backlight illumination (s. fig. \ref{fig.data}). This provides a resolution sufficient to trace the fraction of the larger particles from fig. \ref{fig.size}, but not sufficient to resolve the fraction of smaller particles. 

\section{Results}

\subsection{Data analysis}

\begin{figure*}[h!]
	\centering
	\includegraphics[width=0.99\textwidth]{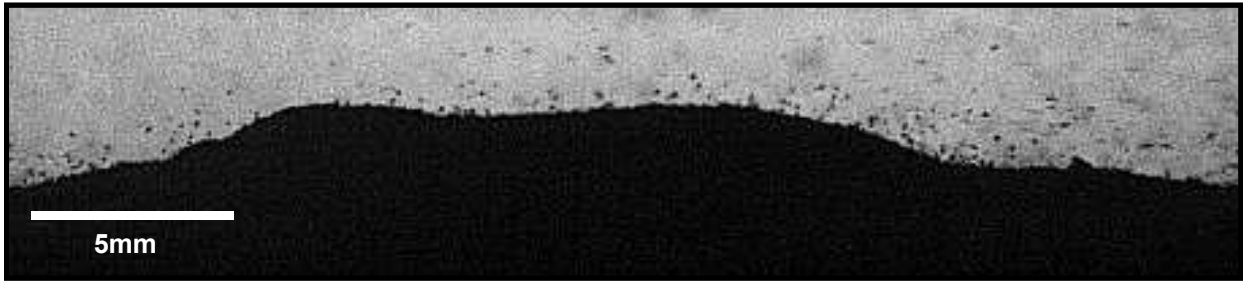}
	\caption{\label{fig.data}Snapshot of particles lifted at 0.38 g close to threshold wind velocity. The wind is flowing from left to right. The shown surface roughness is typical.
	}
\end{figure*}

We use a Martian simulant JSC Mars 1A as soil with a particle density of $1.9$ g/cm$^3$ (or the bulk density of $0.87$ g/cm$^3$ including 54 \% porosity) \citep{Allen1997} and a particle size distribution as depicted in fig. \ref{fig.size}. The shown size distribution represents the volume density of the particle sizes. We cannot exclude that the smaller dust might have an impact on the cohesion properties of the sample. Nonetheless, while the smaller particles get sustained in the atmosphere more easily, saltation is probably dominated by the fraction of the larger particles. The larger particles might have either a grain-like or aggregate structure. In general, they fit in size to particles in Martian dunes, which are given to $87 \mu$m \citep{Claudin2006, Kok2012}. Though even larger particles, e.g. $40-400 \mu$m (High Dune Samples) or $50-400 \mu$m (Namib Dune Sample) are discussed in the literature as well \citep{Ehlmann2017,Tirsch2012,Sullivan2008,Edgett1991} the sample allows an estimation for the minimum shear velocity needed to lift particles. An example for the observation of lifted sand particles at 0.38 g is shown in fig. \ref{fig.data}.

The roughness of the surface is consistent with the roughness map of \citet{Hebrard2012} derived from MOLA data, in which the mean surface roughness on Mars is 4.435 mm and the median surface roughness is 11.05 mm, with 36 \% of the Martian surface having a roughness value higher than 5 mm. The gas flow is just set high enough for lifting events to occur and the fluid threshold shear velocity $u^*$ is determined. Saltation takes place as well as suspension. Once initiated, a lower wind velocity at the impact threshold is needed to sustain the particle flow, but this is not further investigated in this work.

\begin{figure}[h!]
	\centering
	\includegraphics[width=0.49\textwidth]{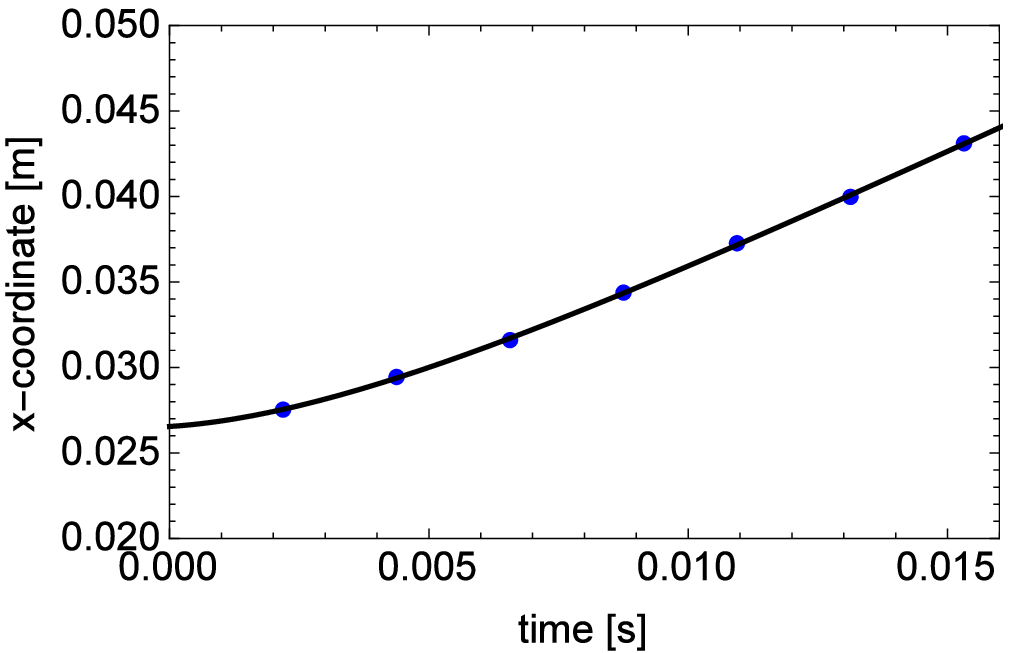}
	\includegraphics[width=0.49\textwidth]{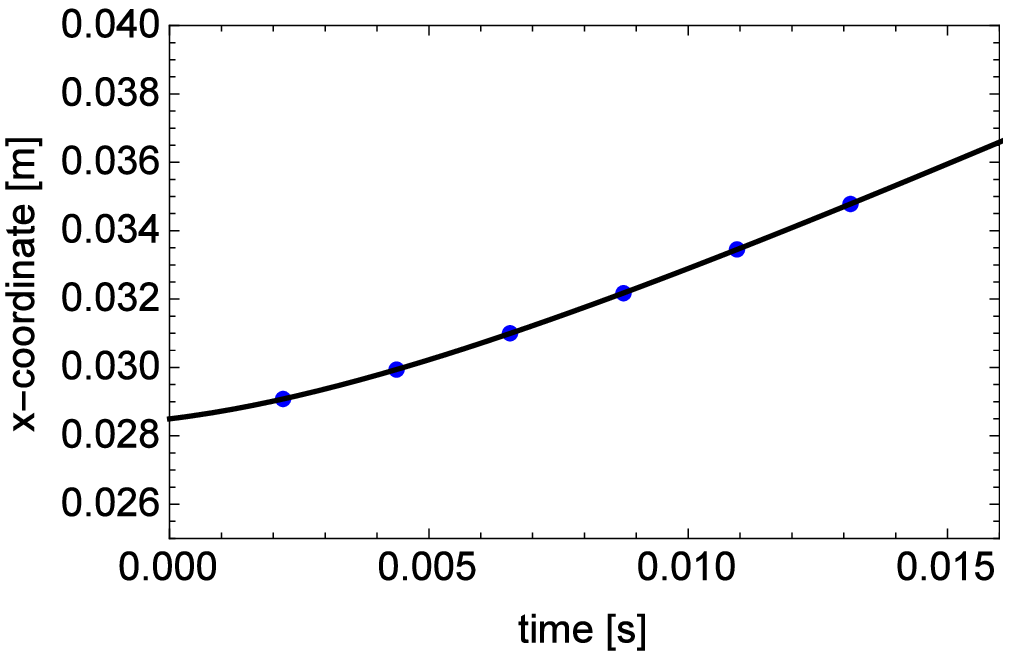}
	\caption{\label{fig.strack}Sample trajectories in wind direction for 1 g (top) and 0.38 g (bottom). The motion of the particles was fitted according to eq. (\ref{eq.trajectory}). The fits are overplotted in black. The particles are accelerated by the gas motion until they finally couple to it.}
\end{figure}

For Martian gravity of 0.38g, 51 trajectories of lifted sand particles are analyzed, while 53 trajectories are analyzed for 1g. From these trajectories, the horizontal gas velocity and its dependency on the height above the sand are calculated. The eroded sand particles couple to the motion of the gas inside the wind tunnel and are used to trace the gas velocity close to the sand bed. For a given height $z$, the trajectory of the sand aggregates along the (horizontal) x-axis can be described by \citep{Wurm2001}
\begin{strip}
\begin{equation}
x(t,z)=\left(v_g(z)-v_0\right)t_C\exp\left(-\frac{t}{t_C}\right)+v_g(z)t+c.
\label{eq.trajectory}
\end{equation}
\end{strip}
This equation is valid for spherical particles with a constant coupling time but can be used as an approximation for bumpy particles as shown in fig. \ref{fig.strack}. The following fit parameters are obtained from fitting the trajectories of the sand particles according to eq. (\ref{eq.trajectory}): The initial velocity $v_0$ of the grain at a certain height $z$, the gas-aggregate coupling time $t_C$, a constant $c$ and finally the gas velocity $v_g(z)$ for a given height $z$ above the dust sample. Furthermore, $t$ is the time after the lifting event. Note, that the Coriolis force is negligible for the lifting process of the particles (as they are at rest) as well as for the grain motion at a constant height $z$ at which the particles are tracked. 

\begin{figure}[h!]
	\centering
	\includegraphics[width=0.49\textwidth]{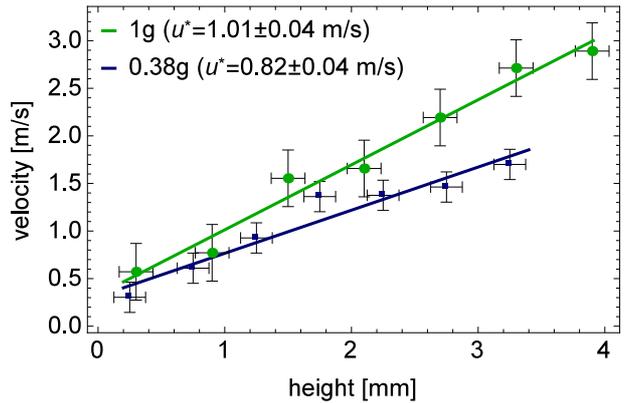}
	\caption{\label{fig.velocity}Gas velocity profile over height at the threshold of particle lifting for 1 g (green) and 0.38 g (blue). The data are binned, including 53 individual values for 1 g and 51 values for 0.38 g. The slopes $dv_g(z)/dz$ resulting from the linear fits are 680 $s^{-1}$ for 1 g and 453 $s^{-1}$ for 0.38 g. The threshold $u^*$ was calculated using eq. (\ref{eq.shearvelocity}).}
\end{figure}

For the 0.38g trajectories, the values for the gas velocities are binned in 0.5 mm steps. For each bin, the values for the median gas velocity are calculated with 7-8 individual values for the gas velocity from the fitted trajectories according to eq. (\ref{eq.trajectory}). For the 1g trajectories, the bin size is set to 0.6 mm. The binned data is given in fig. \ref{fig.velocity}. 
Both profiles indicate a linear correlation between the horizontal gas velocity and the height above the sand surface. A linear profile close to the ground is also in agreement with former experiments in wind tunnels \citep{Merrison2008}. The error bars show the mean error calculated from the fits of all trajectories for 0.38 g and 1 g, respectively.

For the microgravity measurements, additional uncertainties resulting from the residual acceleration and vibrations inside the aircraft have to be considered. In order to avoid errors due to the residual acceleration of $\pm$0.05 g, the gas velocities deduced from the trajectories are averaged as described. The vibrations inside the cabin have a higher frequency of $\omega\approx$10 Hz, which is an estimation from the acceleration data. If the amplitude of the vibrations was $A\approx 100 \mu$m, which equals the grain size and hence is a maximum estimation, the error in acceleration would be on the order of $g_v= A\omega^2\approx 0.01 m/s^2$. Compared to the Martian gravitation of 0.38g, this is a relative error of 2.7\%, which we consider as negligible.

\subsection{Threshold shear velocity and cohesion reduction for Mars}

Interaction of a turbulent wind flow with a surface can be characterized by the shear velocity  $u^*  = \sqrt{\tau/\rho}$ with the shear stress $\tau$ and the fluid density $\rho$ \citep{Schlichting2016}.  This quantity can be interpreted as the wind velocity acting directly at the soil. 
The shear stress can be expressed by Newton's law of viscosity to $\tau = \eta dv_g(z)/dz$ with the dynamic viscosity $\eta$ and the flow height profile $dv_g(z)/dz$ depending on the gas velocity $v_g(z)$ and the height $z$. Thus, $u^*$ can also be expressed as
\begin{equation}
u^* = \sqrt{\frac{\eta}{\rho}\frac{dv_g(z)}{dz}}.
\label{eq.shearvelocity}
\end{equation}
The gas velocity $v_g(z)$ is logarithmic in $z$ within a turbulent sublayer and linear in $z$ within a viscous sublayer close to the ground as measured in this work. Considering $\eta\approx15$ $\mu$Pa$\cdot$s and $\rho\approx0.01$ kg/m$^3$ (CO$_2$ at 6 mbar and 300 K) as well as $dv_g(z)/dz$ from fig. \ref{fig.velocity}, the threshold shear velocity can be derived directly from eq.(\ref{eq.shearvelocity}) and yields $0.82\pm0.04$ m/s for 0.38 g and $1.01\pm0.04$ m/s for 1 g. 

The threshold shear velocity at 0.38 g is lower than values determined in prior experiments on ground \citep{Greeley1980, Merrison2008} which are generally somewhat larger with $\sim 1.5-2$ m/s. However, $u^*$ was measured in a different gravitational environment in this work and depends also on the grain species. Thus, it cannot be compared directly to these other works. This might also be an indication that prior experiments perhaps overestimated this value for the Martian soil. 

Using the models from \citet{Shao2000} and \citet{Merrison2008} with the threshold shear velocities for 0.38 g and 1 g, the particle density of $1.9$ g/cm$^3$ and a mean particle diameter of approximately 85 $\mu$m we get a surface energy of $\gamma_{SL} \approx 1.1 \cdot 10^{-7}$ J/m$^2$. This is an unreasonably low value as the used JSC species mostly consists of $SiO_2$, $Al_2O_3$, $Fe_2O_3$ and $CaO$ which all exceed values of $10^{-2}$ J/m$^2$ for the surface energy \citep{Heim1999,Miller2011}. 

In consequence of the low value for $\gamma_{SL}$ we consider a lower cohesive force at lower gravity influencing the ratio of the determined threshold shear velocities. The cohesion force at the threshold can be estimated from the force balance
\begin{equation}
C_L\frac{\pi}{2}\rho r^2u^{*2} = \sum_j F_{C,j}+Mg.
\label{eq.forcebalance}
\end{equation}
The lifting force is given on the left side \citep{Kuepper2015}. $C_L$ is the lifting coefficient which depends on the boundary conditions of the wind tunnel and the shape of the particles, $r$ is the average radius of the particles and $\rho$ is the fluid density. Counteracting are the gravitational force $M \cdot g$ with the particle's mass $M$ and the gravitational acceleration $g$ and the sum over all cohesive contacts  $F_{C,j}$ of a grain. Grains of $\sim$100 $\mu$m are usually easiest to move as cohesive forces and gravity are similar \citep{Greeley1980}. Thus, none of the addends can be neglected. We assume that all individual contacts are sharing the same contact area and can be described by the JKR model \citep{Johnson1971,Tomas2006} which gives $\sum_jF_{C,j}\approx NF_{J}=N \frac{3}{2}\pi\gamma r$ with the amount of contacts per grain $N$ and the surface energy $\gamma$. This approach finally results in the threshold
\begin{equation}
 u^* = \sqrt{\frac{4}{3 C_L}\left(\frac{9N}{2}\frac{\gamma}{\rho d}+\frac{\rho_p}{\rho}dg\right)},
\end{equation}
with the particle density $\rho_p$ and diameter $d$. Except the dependency in $N$, this expression is similar to the equation provided by \citet{Shao2000}. If the contact number $N$ depends on the gravitational acceleration, $u^*$ might be lower for reduced gravity. With two values for the $u^*$, this dependency $N(g)$ can be estimated.

The ratio between both threshold shear velocities at different gravitational environments can be written as
\begin{equation}
\frac{u_1^{*2}}{u_2^{*2}}=\frac{N_1F_{J}+Mg_1}{N_2F_{J}+Mg_2}\equiv\frac{F_N+Mg_1}{\chi F_N + Mg_2},
\label{eq.udivu}
\end{equation}
with the sum of all contact forces $F_N\equiv N_1F_{J}$ and the contact number ratio $\chi\equiv N_2/N_1$. We can derive $\chi$ from eq. (\ref{eq.udivu}) to
\begin{equation}
\chi=\frac{u_2^{*2}}{u_1^{*2}}\left(1+\frac{Mg_1}{F_N}\right)-\frac{Mg_2}{F_N}.
\label{eq.contactratio}
\end{equation}
Applying the values for the fluid threshold shear velocity in this work with the average number of contacts $N_1$ in 0.38g and $N_2$ in 1g gives
\begin{equation}
\chi\approx\frac{3}{2} \hspace{3mm} \forall \hspace{3mm}  F_N \gg 10^{-8}N.
\label{eq.conred}
\end{equation}
This result shows, that the average number of contacts and thus also the total contact forces are only 2/3 as large in 0.38 g as in 1 g, if $F_N$ exceeds $ 10^{-8}$ N by an order of magnitude. If we consider $N=1$, $\gamma \approx 0.01$ J/m$^2$ which is a typical value for silicate spheres \citep{Heim1999} and $r=10^{-5}$ m (as minimum estimation) the additional condition is easily fulfilled with $F_1=\frac{3}{2}\pi\gamma r\approx 5\cdot10^{-7}$ N. Experimental work on contact forces confirms this likewise \citep{Heim1999}.
This is the first time that it is considered that cohesion is not constant in soils
of different planets as gravity does compress the soil differently. A reduction in contact number in the low gravity environment of Mars
can explain a reduction in the threshold wind velocity necessary to lift particles.
Absolute values of the fluid threshold shear velocity derived from our experiment under 0.38 g indicate that saltation and suspension are possible under the conditions given on Mars and in agreement to particles being observed in motion.

\section{Simulation with the Global Circulation Model (GCM)}

\subsection{Mars GCM}

A General Circulation Model (GCM) for the atmosphere of Mars is applied to calculate surface shear velocities \citep{Daerden2015,Neary2018}. It is operated on a grid with a horizontal resolution of 4$^\circ$x4$^\circ$ and with 103 vertical levels reaching from the surface to $\sim$150 km. The model calculates heating and cooling of atmospheric CO$_2$ and dust and ice particles by solar and IR radiation and solves the primitive equations of atmospheric dynamics. The geophysical boundary conditions are taken from observations and include a detailed surface roughness length map. Physical parameterizations in the model include an interactive CO$_2$ condensation and surface pressure cycle, a thermal soil model, turbulent transport in the atmospheric surface layer and convective transport inside the planetary boundary layer. The effects of the extreme Martian topography are considered with a low level blocking scheme. The shear velocity is derived from the computed wind field in the second lowest vertical model level (at height $\sim$15 m), following the expressions derived from similarity theory \citep{Jacobson2005}. In the model, dust is lifted by saltation whenever the shear stress exceeds a critical value that is calculated from the threshold shear velocity given by eq. (\ref{eq.shearvelocity}), where $dv_g/dz$ is taken from fig. \ref{fig.velocity}, and the dynamic viscosity for CO$_2$ is calculated from Sutherlands formula after \citet{Crane1988} using the GCM predicted surface temperature. Dust is lifted in a lognormal distribution with mean radius 1.5 $\mu$m, which contains 3 size bins: 0.1, 1.5 and 10 $\mu$m. The idea of saltation is that the larger sand particles are lifted, if the shear velocity is exceeded and fall back to the surface as they are too large to stay aloft. From the collision with the surface, smaller (dust) particles are lifted, which are able to go in suspension. In the GCM model this process is shortcut by lifting $\mu$m-size particles directly when the threshold shear velocity is exceeded.

Dust is lifted in the GCM following the Kahre-Murphy-Haberle (KMH) method \citep{Kahre2006}, in which the dust mass flux from the surface is calculated as 
\begin{equation}
F = \left(2.3 \cdot 10^{-3}\right) \alpha \tau^2\left(\frac{\tau-\tau^*}{\tau^*}\right)
\end{equation}

with $\tau$ the actual and $\tau^*$ the threshold surface wind stress. $\alpha$ is a proportionality factor that has to be set for an optimal match with observations. It does not control where and when dust is lifted, but only how much dust is actually lifted. Dust particles are sedimented in the model using the size-dependent Stokes settling velocity with Cunningham slip-flow correction \citep{Jacobson2005}. Dust is radiatively active in the GCM, by a 2-stream approximation applying the latest optical properties  \citep{Wolff2006,Wolff2009}. It undergoes all the transport processes in the model such as diffusive mixing and advection. Dust is the main thermodynamic agent in the middle and lower atmosphere of Mars and drives the global circulation under differential solar heating in combination with local processes such as saltation. In this way, saltation is simulated in the GCM as fully interactive. One assumption that is made is that of a limitless surface reservoir of dust. 

Until now, in GCMs a threshold of typically 0.0225 Pa is used \citep{Haberle2003,Kahre2006,Daerden2015,Neary2018}, a value corresponding to a $\sim$40\% reduction of the critical shear stress derived from the original lab measurement for static conditions \citep{Greeley1980}, to have any dust lifting at all. It is found in our simulations that the new threshold shear stress can be typically 5 times lower than the one derived from previous laboratory data, and so that the new threshold for saltation on Mars does no longer require GCMs to apply a strongly reduced value to simulate dust lifting.

\subsection{Simulations with the new threshold values}

\begin{figure*}[h!]
	\centering
	\includegraphics[width=0.99\textwidth]{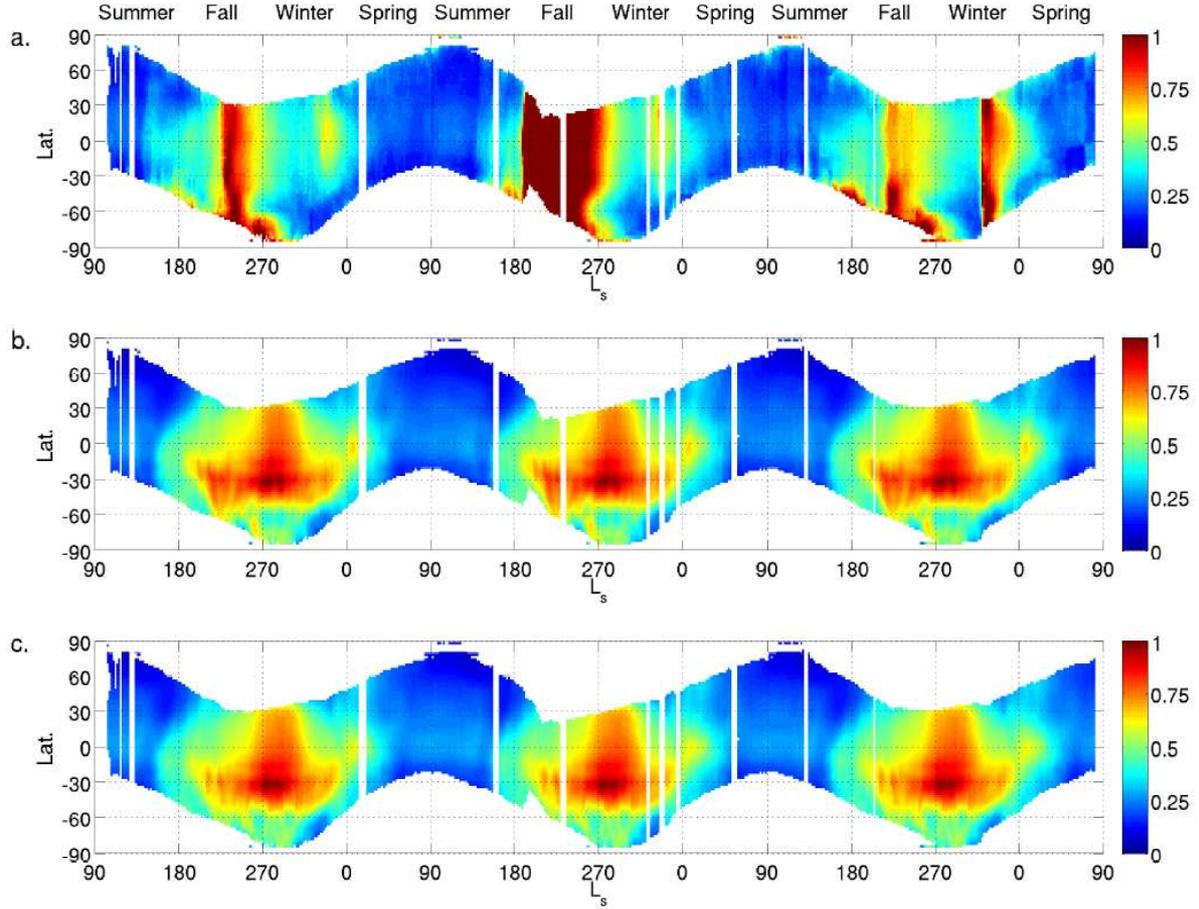}
	\caption{\label{fig.GEM}(a) Latitude versus time distribution of the dust optical depth measured on Mars by the TES instrument on MGS for 3 consecutive Mars years (1999-2004). The horizontal coordinate is the solar longitude ($L_S$). The data was scaled to visible wavelengths and to a surface pressure value of 610 Pa, and binned over 2$^\circ$ latitude and 2$^\circ$ $L_S$. (b) Simulation of the same quantity in the GCM by applying a threshold on surface shear velocity that was strongly reduced from previous experimental work. (c) Simulation of the same quantity in the GCM by applying the threshold on surface shear velocity derived from the experiment in this work, without reduction. Model output is averaged in the same way as the data and removed where no data is available. 		
	}
\end{figure*}

Using the new values for the threshold shear velocity with the GCM allows a prediction of locations where dust and sand movement is preferred on the Martian surface. Fig. \ref{fig.GEM}a shows the zonally averaged dust optical depth measurements from the Thermal Emission Spectrometer (TES) instrument on the NASA Mars Global Surveyor (MGS) orbiter \citep{Smith2004} for 3 consecutive Martian years. The seasonal behavior with a less active dust season during northern spring and summer and a highly active dust (or dust storm) season during southern spring and summer is clearly visible. 

The dust data from TES shown in fig. 6a are extinction optical depths obtained from the measured absorption optical depths by multiplying by 1.3 \citep{Smith2004}. The optical depths are scaled to visible wavelengths from the original measurement at 1075 cm$^{−1}$ (9.3 $\mu$m) by multiplying by 1.8 \citep{Clancy2003}. The TES measurements are mostly taken around 2 p.m. local time \citep{Smith2004}. The values are scaled to a surface pressure of 610 Pa and are averaged over all longitudes and over bins of 2$^\circ$ in latitude and 2$^\circ$ in solar longitude. Solar longitude ($L_S$) is the angle from the sun between Mars and its orbital vernal equinox and often used to indicate Mars time of year.

\ref{fig.GEM}b shows the result of a simulation of the dust cycle with the GCM applying the old threshold for dust lifting that was strongly reduced from previous experimental work \citep{Haberle2003}. The value of the efficiency factor $\alpha$ was set to 0.015. \ref{fig.GEM}c shows the result of a simulation of the dust cycle with the GCM applying the new threshold for dust lifting derived from our experiment, without any further reduction. The efficiency factor $\alpha$ was set to 0.0026. The model results presented in the figure are obtained as follows from the GCM output. The dust optical depth is calculated in the model at 0.67 $\mu$m. The model output is sampled every 30 minutes, and averaged over all longitudes with local time between 1 and 3 p.m. The resulting dataset is binned like the TES data over 2$^\circ$ in $L_S$. A mask was applied to the resulting time series to remove the times and latitudes for which there is no TES data available.

The figure shows that the GCM is able to predict the times and latitudes where dust lifting occurs and provides a dust cycle that is qualitatively comparable to the data. The interannual variability of the peaks in the dust storm season is a topic of ongoing research \citep{Mulholland2013,Shirley2017} and beyond the scope of the present work. The figure also shows that applying the new threshold obtained from our measurements in the model simulation is equivalent to applying the threshold that was strongly reduced from values obtained in previous measurements, i.e. the new threshold does not lead to unforeseen complications, and allows for a dust cycle simulation using an experimentally found threshold for saltation.

\section{Discussion and Conclusion}

We measured the fluid threshold shear velocity for a Martian simulant JSC 1A with a dominating grain size on the order of $\sim$ 100 $\mu$m. For Martian gravity of 0.38 g, this value yields $0.82\pm0.04$m/s and increases to $1.01\pm0.04$ m/s for 1g. 
We attribute the difference between both threshold shear velocities to a reduced number of contacts between particles. Under Martian gravity the number of contacts is reduced by a factor of 2/3 in comparison to Earth's gravity. In the wind tunnel experiments, this is established as the sand bed is disturbed at the beginning of each parabola. The top layers of the sand bed are lifted by the shutter mechanism and reset themselves in low gravity. The top layers therefore represent the mechanical structure of a sand bed prepared under Martian gravity. The results derived from eq. (\ref{eq.conred}) are an approximation for a continuum flow. For the case of a slip flow, the drag force is proportional to the shear velocity and has to be corrected with an additional Cunningham correction factor \citep{Kuepper2015}. In this case, the reduction of the contact number would rather be described by
\begin{strip}
\begin{equation}
\chi=\frac{u_2^{*}}{u_1^{*}}\left(1+\frac{Mg_1}{F_N}\right)-\frac{Mg_2}{F_N} \approx 1.22\, \hspace{3mm} \forall \hspace{3mm}  F_N \gg 10^{-8}N. 
\end{equation}
\end{strip}

The low absolute value for the threshold shear velocities also shows the importance of the chosen sand and the conditions for the sample preparation. Prior experiments  perhaps overestimated the fluid threshold shear velocity on Mars, as the soil sample was always prepared under Earth gravity and therefore with the corresponding number of contacts between the particles. Our new findings bring numerical simulations of dust transport on Mars by general circulation models in agreement with observations, without the need for reduction of the threshold.

With this work, we perform the first wind tunnel studies on saltation directly under Martian gravitational and atmospherical conditions. Nonetheless, the experiments are performed on a small time scale of $\sim$20 s. It is an important question whether the results would be applicable on longer timescales which cannot be answered by this work. In future, further quantitative studies comprising experiments for several g-levels might confirm the tendencies and give a clearer picture of the relation between gravity and cohesion.

\appendix
\textbf{Acknowledgements}
The experiments were carried out on the 65th ESA parabolic flight campaign as part of the Fly Your Thesis! 2016 programme. The work was supported by ESA Education, the DFG under grant number WU 321/12-1 and DLR Space Administration with funds provided by the Federal Ministry for Economic Affairs and Energy (BMWi) under grant number DLR 50 WM 1760. We thank Jan Raack and an anonymous reviewer for a constructive review.


\end{document}